\documentstyle[12pt,psfig,epsfig,aaspp4]{article}

\def\ergcmsec{\hbox{erg cm$^{-2}$ s$^{-1}$}}

\def\degmark{^\circ}
\def \rsun {\ifmmode$R$_{\odot}\else R$_{\odot}$\fi}

\def \hcm {\hbox {\ifmmode $ H atoms cm$^{-2}\else H atoms cm$^{-2}$\fi}}
\def\approxgt{\mathrel{\hbox{\rlap{\lower.55ex \hbox {$\sim$}}
        \kern-.3em \raise.4ex \hbox{$>$}}}}
\def\approxlt{\mathrel{\hbox{\rlap{\lower.55ex \hbox {$\sim$}}
        \kern-.3em \raise.4ex \hbox{$<$}}}}

\newcommand {\ROSAT} {{\it ROSAT}}

\newcommand {\asca} {{\it ASCA}}
\newcommand {\sax} {{\it BeppoSAX}}

\lefthead{Owens et al.}
\righthead{X-ray Emission from Comet Hale-Bopp}

\begin{document}

\title{Evidence for Dust Related X-ray Emission from Comet C/1995~O1 
(Hale$-$Bopp)\altaffilmark{1}}

\author{Alan Owens, A. N. Parmar, T. Oosterbroek, A. Orr}
\affil{Astrophysics Division, Space Science Department of ESA,
ESTEC, 2200 AG Noordwijk, The Netherlands}

\author{L. A. Antonelli, F. Fiore\altaffilmark{2}}
\affil{\sax\ Science Data Center, Nuova Telespazio, 
via Corcolle 19, I-00131 Roma, Italy}

\author{R. Schulz}
\affil{Solar System Division, Space Science Department of ESA, 
ESTEC, 2200 AG Noordwijk, The Netherlands}

\author{G. P. Tozzi}
\affil{Osservatorio Astrofisico di Arcetri, Largo E. Fermi 5, 
I-50125, Firenze, Italy}

\author{M. C. Maccarone}
\affil{IFCAI/CNR, via U. La Malfa 153, I-90146, Palermo, Italy}

\and

\author{L. Piro}
\affil{IAS/CNR, via Enrico Fermi 21-23, I-00044, Frascati, Italy}

\altaffiltext{1}{Partially based on observations carried out at the European
Southern Observatory (ESO), La Silla, Chile}

\altaffiltext{2}{and Osservatorio Astronomico di Roma, I-00044, Monteporzio,
Catone, Italy}

\authoraddr{Astrophysics Division, Space Science Department of ESA,
Postbus 299, 2200 AG Noordwijk, The Netherlands}

\begin{abstract}
We report the discovery of X-ray emission from comet 
C/1995 O1 (Hale-Bopp) by the LECS instrument on-board \sax\ on 1996 September 
10--11. The 0.1--2.0~keV luminosity decayed by a 
factor of 2 on a timescale of $\sim$10 hr 
with a mean value of $5 \times 10^{16}$~erg~s$^{-1}$.
The spectrum is well fit by a thermal bremsstrahlung 
model with a temperature of $0.29 \pm 0.06$~keV, or a power-law
with a photon index of $3.1 \pm ^{0.6}_{0.2}$. 
The lack of detected C and O line emission places severe
constraints on many models for cometary X-ray emission, especially those 
which involve X-ray production in cometary gas.
The luminosity is a factor of at least 3.4 greater than measured by 
{\it Extreme Ultraviolet Explorer} ({\it EUVE}) 4 days later. 
This difference may be related to the emergence from the nucleus 
on 1996 September 9 of a dust-rich cloud. 
Over the next few days the cloud continued to expand becoming 
increasingly tenuous, until it had reached an extent of 
$\sim$3 $\times$ 10$^5$~km (or $\sim$2$'$) by the start of 
{\it EUVE} observation. 
We speculate that the observed reduction in X-ray 
intensity is evidence for dust fragmentation. 
These results support the view that cometary X-ray 
emission arises from the interaction between solar X-rays
and cometary dust.
\end{abstract}

\keywords{comets: general --- comets: individual (Hale-Bopp)
--- X-rays: general}

\section{INTRODUCTION}

The \ROSAT\ observations of X-ray emission from comets C/1996 B2 (Hyakutake) 
and C/1990 N1 (Tsuchiya-Kiuchi) came as a considerable surprise (Lisse et 
al. 1996; Dennerl et al. 1996b) and despite considerable effort there is 
still no generally accepted model for cometary X-ray emission. 
The emissions were centered a few arc-minutes ($\sim$2 $\times$ 10$^{4}$ km) 
from the nucleus and had an extent of 5--15\arcmin, or (3--8) $\times$ 
10$^{4}$~km, being elongated normal to the Sun-nucleus line. 
The observed fluxes are a factor of $\sim$10$^3$ greater than predicted 
by early models in which X-rays 
are generated by fluorescent emission and scattering of solar X-rays in the 
coma (see Krasnopolsky (1997) for a review). 
Models which generate X-rays in the coma via solar 
wind proton, or electron, interactions also suffer from low efficiencies. 
In view of these difficulties, several models have emerged which
predict higher cometary X-ray intensities (see Table~1). In the 
solar wind models of H\"aberli et al. (1997) and Cravens (1997) X-rays 
are generated following charge exchange excitations of highly ionized 
solar wind ions with neutral 
molecules in the comet's atmosphere. Bingham et al. (1997) propose that
cometary X-rays are produced by energetic electrons 
generated by plasma wave turbulence. The turbulence is assumed to result 
from the relative motion of the cometary plasma and the solar wind.
The attogram dust models of Wickramasinghe \& Hoyle (1996) and 
Krasnopolsky (1997) produce X-rays from the scattering, fluorescence 
and bremsstrahlung of solar X-rays in attogram ($\sim$10$^{-18}$~g) dust 
particles, such as those detected in the wake of comet Halley 
(Utterback \& Kissel 1990). Recently sub-micron sized grains
have been observed in the coma of Hale-Bopp (Williams et al. 1997).
 
At the time of the \sax\ observation, 
Hale-Bopp (C/1995 01; Hale \& Bopp 1995) was 2.87 AU from the 
Earth at a heliocentric distance of 3.13~AU and at a relatively 
constant brightness of m${\rm _v \sim 6}$. The Sun was at a 
Position Angle of 274$\degmark$ and the comet phase angle 
(Sun-comet-Earth angle) was 18\fdg7.
The size of the coma was estimated to be $\sim$20\arcmin\ 
($\sim$2.5 $\times$ 10$^{6}$ km)
and some observers reported a short dust tail of up to 1\fdg5 
(10$^{7}$~km). The 
diameter of the nucleus is estimated by the {\it Hubble Space Telescope} to 
be 27--42~km (Weaver et al. 1997).
In 1996 September Hale-Bopp was observed by the {\it Extreme Ultraviolet
Explorer} ({\it EUVE}), \ROSAT, \asca, and \sax\ satellites. We present
here results from the \sax\ observation briefly reported in Owens
et al. (1997). Results from {\it ASCA} can be found in Kellett et al. (1997) and 
{\it EUVE} in Mumma et al. (1997) and Krasnopolsky et al. (1997a).

\section{X-RAY OBSERVATION}

The Low Energy Concentrator Spectrometer (LECS, 0.1-10~keV, 
Parmar et al. 1997) is one of five instruments 
on-board \sax\ (Boella et al. 1997a). It 
comprises an imaging mirror system
and a driftless gas scintillation proportional counter. 
The energy resolution is 32\% at 0.28~keV, the field of view (FOV) 
is 37\arcmin\ diameter (4.6 $\times$ 10$^{6}$ km at the comet) 
and the full width at half-maximum (FWHM) of the encircled energy 
function is $11 \farcm 5$ at 0.2~keV, 
falling to 5\farcm1 at 1~keV. The effective area just below the 
C edge at 0.28~keV is 20~cm$^2$. The in-orbit background counting rate is 
9.7 $\times$ 10$^{-6}$ arcmin$^{-2}$ s$^{-1}$ keV$^{-1}$.
\sax\ was launched into a 600 km equatorial 
orbit on 1996 April 30.

Hale-Bopp was observed by \sax\ between 1996 September~10 04:55 
and September~11 04:21~UTC. The LECS exposure is 11.5~ks. A weak
(2.1 $\pm$ 0.3) $\times$ 10$^{-12}$~\ergcmsec; 0.1--2.0~keV) source 
is visible 
3\farcm5 off-axis. The source position is $<$2\arcmin\ ($<$2.6  
$\times$ 10$^{5}$ km) from the comet's
nucleus (see below) and examination of the \ROSAT\ All Sky Survey catalog 
(Voges et al. 1996) reveals no known X-ray sources at this position.
Using the \ROSAT\ logN-logS relation of Hasinger et al. (1993) the 
probability of randomly detecting an X-ray source as bright as this
within the extraction region is $6 \times 10^{-3}$. We exclude
a UV leak as being responsible for the counts since the LECS has
made a number of deep observations of UV bright stars such as Polaris
and Capella and no unexpected features were found.
In addition, there are no unusual objects in the SIMBAD database and
inspection of several deep LECS exposures revealed no ``hot spots'', or
other instrumental features at this position. 
The Medium Energy Concentrator Spectrometer 
(MECS; 1.3--10~keV; Boella et al. 1997b) detected excess 1.3--2.0~keV
emission at a position consistent with the LECS detection (at 
1.3$\sigma$ confidence) with an intensity
of $(3.3 \pm 2.5) \times$ 10$^{-14}$~erg~cm$^{-2}$~s$^{-1}$.
This is in good agreement with LECS 1.3--2.0~keV intensity of 
$(1.8 \pm 1.1) \times$ 10$^{-14}$~erg~cm$^{-2}$~s$^{-1}$.
Based on the above, we identify
comet Hale-Bopp as the source of the detected X-rays.

A motion corrected image of the 
region of sky containing Hale-Bopp is shown in Fig.~1. 
The image is re-binned to a 56\arcsec\ $\times$ 56\arcsec\ pixel 
size and smoothed  using a Gaussian filter of width 1\farcm5 
(1.9 $\times$ 10$^5$~km).
The direction of the comet's motion and the position of the Sun are 
indicated. 
The bulk of the emission originates on the sunward side of the 
comet in agreement with \ROSAT\ images of comet Hyakutake (Lisse et al. 1996). 
The extent of the emission is consistent with the LECS point spread 
function (PSF) of 9\farcm5 FWHM at the mean energy of the detected emission, 
although 
the width normal to the Sun-nucleus axis appears wider than in 
the direction of motion, similar to that observed in other comets. 
The 68\% confidence limit to any source extent is $<$6\farcm5 or
$<$$8.1 \times 10^5$~km.
At the start of the LECS observation, the source position 
is RA = 17$^{\rm h}$ 33$^{\rm m}$ 43\fs4,
decl. = $-$6$^{\circ}$ 01\arcmin\ 12\arcsec\ (J2000) with a 68\% confidence
uncertainty radius of 1\farcm0. This is $1\farcm7 \pm 1\farcm0$, 
or $(2.1 \pm 1.3) \times 10^5$~km, from the position
of the nucleus, in the general solar direction.

The source spectrum was extracted from within an 8\arcmin\ radius of 
the mean source position and analyzed using version 1.4.0 of the 
SAXLEDAS data analysis system. No correction for the comets 17\arcsec~hr$^{-1}$
motion was applied since most of the events arrive within the
first $3\times 10^4$~s (see Fig.~2) when the movement
of $<$2\farcm5 is small compared to the size of the 
extraction region. The total number of counts is 246, whereas 
113 are obtained at the same position in the standard background exposure. 
The comet's position at the time of the LECS observation was 
near the southern tip of the North Polar Spur. The low-energy 
X-ray background in 
this area is higher than in the areas used in the standard LECS 
background field ({\it e.g.}, Snowden et al. 1995). For this reason  
the background spectrum was extracted from the image itself 
using an 8\arcmin\ radius region centered 16\arcmin\ diametrically opposite 
in the LECS FOV. A small correction for mirror vignetting was applied.
After background subtraction, the count rate is 
$0.0120 \pm 0.0018$~s$^{-1}$.  

The extracted spectrum is equally well fit (with $\chi^2$'s of 
9 for 9 degrees of freedom (dof)) by a thermal bremsstrahlung 
model of temperature $0.29 \pm 0.06 $~keV or a 
power-law model of photon index $3.1 \pm ^{0.6}_{0.2}$ 
(Fig.~3). 
A blackbody gives a less acceptable fit with a $\chi^2$ of 18 for 9 dof
(P($>$$\chi^2$)=3\%).
The 0.1--2.0~keV spectrum predicted by the Solar wind model 
of H\"aberli et al. 
(1997) was also fit to the LECS data. This spectrum consists of a 
series of narrow lines resulting from the excitation of highly charged 
states of O, 
C, and Ne ions (see Table~2 of H\"aberli et al. 1997). The line species 
are not expected to 
vary appreciably from comet to comet, and so the line energies were 
fixed at the values given in H\"aberli et al. (1997). The best-fit 
$\chi^2$ is 44 for 6 dof, with 3 of the lines requiring zero flux. 
The 95\% confidence limits to C and O narrow line emission at
0.28 and 0.53~keV are $1.0 \times 10^{15}$ and 
$7.8 \times 10^{15}$~erg~s$^{-1}$, respectively.
This implies that any such narrow line 
luminosity must be $<$18\% of the 0.1--2.0~keV continuum luminosity. 
 
The 0.1--2.0~keV luminosity is $4.8 \times 10^{16}$~erg~s$^{-1}$, or
$2.1 \times 10^{-12}$~erg~cm$^{-2}$~s$^{-1}$, for the best-fit thermal 
bremsstrahlung spectrum and $5.9 \times 10^{16}$~erg~s$^{-1}$, or
$2.6 \times 10^{-12}$~erg~cm$^{-2}$~s$^{-1}$, for the best-fit power-law 
spectrum. Using the 0.07--0.18~keV {\it EUVE} Deep Survey camera, 
Krasnapolsky et al.
(1997), assuming spectrally uniform emission, derive 0.095--0.165~keV 
fluxes of $8 \times 10^{24}$ and $1.5 \times 10^{25}$~photons~s$^{-1}$ for
radii of $4 \times 10^5$ and $10^6$~km, respectively.
This narrower energy range is used by Krasnapolsky et al. (1997) to
reflect the ``effective'' instrument bandpass.
In addition, the LECS source is point-like (with an extent of 
$<$$8.1 \times 10^5$~km), while that observed by {\it EUVE} is extended.
It is therefore not straightforward to compare LECS and {\it EUVE}
fluxes. Using the {\it EUVE} flux within the upper-limit LECS
source extent, gives a LECS to {\it EUVE} flux ratio of $8.3 \pm 1.4$.
(Note, the {\it EUVE} flux was derived assuming spectrally uniform emission.)
Using the larger source radius reduces this ratio to $4.4 \pm 1.0$.
The above uncertainties include the contribution from the error
on the LECS temperature determination.
Although the measured extent of the dust cloud ($\sim$2--4 $\times~10^5$ km) 
at the time of the {\it EUVE} observation is comparable to the smaller radius, 
we take the worst case LECS to {\it EUVE} flux ratio to be $>$3.4. 

Figure~2 shows light curves for the same source and 
background regions as for the spectrum. The integration interval is 2000~s. 
A best-fit constant count rate to the background region data
yields a $\chi^2$ of 14 for 15 dof, whereas for the source region 
the $\chi^2$ is 44 for 13 dof. Figure~2 indicates
that the X-ray intensity of the comet gradually decreased during the 
observation and is variable by a factor 
of $\sim$2 on a timescale of hours. Modeling the light curve by an 
exponential decay, reduces the $\chi^2$ to 18 for 12 dof for a 1/e time 
of 9.3 hr. 

\section{OPTICAL OBSERVATIONS}

Contemporary optical observations were carried out on the 2.2~m telescope
at La Silla using the multi-mode EFOSC~II instrument in 3 narrow bands
to differentiate between dust ($\lambda$=4467~\AA~, $\Delta \lambda$=56~\AA~FWHM) 
and C$_2$ ($\lambda$=5117~\AA~, $\Delta \lambda$=56~\AA~FWHM) 
and CN ($\lambda$=3849~\AA~, $\Delta \lambda$=87~\AA~FWHM) gas emissions. Hale-Bopp 
was observed at about 01:00 UTC on the nights of 1996 September 10
to September~17 with exposure times of 180~s (dust), 180~s (C$_2$)
and 300~s (CN). Each image was divided by the previous one so that temporal 
changes in the coma become apparent. This technique removes the strong smooth 
intensity profile of the coma so that superimposed time variable structures 
become visible (Schulz 1991).

The quotient images (Fig. 4, top) and differential brightness profiles 
(the lower panel of Fig. 4) show 
that a dust cloud was ejected from the 
nucleus which expanded outwards
and appeared as a shell when integrated over the line of sight. 
The original images show that the bulk of the dust was emitted towards the 
North and North-West. Additionally, four jets are apparent 
in Fig.~4$a$ confirming the 
observations of Kidger et al. (1996). They are suppressed in subsequent 
images because the exposures were taken at intervals 
of very nearly twice the rotation period ($\sim$11.5~hr) of the nucleus. 
The jets are unlikely to be related to the cloud as multiple jets were 
continually observed during 1996 August and September (Kidger et al. 1996).

The transverse dust velocities were determined by measuring the position 
of maximum emission in radial brightness profiles derived from the original 
images (see the lower panel of Fig. 4). By extrapolating backwards, the onset of the outburst 
occurred
on 1996 September 9 ($9.4 \pm 1)$~UTC. The thickness of the shell is a function 
of the outburst lifetime and the velocity distribution of the dust. The images 
indicate that the outburst lasted for $\le$1~day, after which dust ejection 
ceased and the 
shell expanded outwards at an average velocity of 0.11 km~s$^{-1}$, 
becoming increasingly tenuous. 
The temporal evolution of the dust radial intensity profiles is 
shown in the lower panel of Fig. 4. The quiescent intensity is taken to 
be that measured 2~$\times$
10$^5$ km from the comet on 1996 September 10. This distance is a factor of 
$\sim$10 times further than the outburst dust had traveled, and so provides an
uncontaminated measure. Assuming the dust column density is proportional to 
its optical intensity, 
the ratio of maximum to quiescent intensities gives an estimate of
the change in dust column density. This is at least a factor of 7 
greater at the time
of the LECS
observation. 
After September 10, the dust cloud density decreased at an initial rate 
of $\simeq$40\% per day as the shell expanded outwards.
On September 11 the cloud had an angular extent of $\sim$0\farcm8 or 
$\sim$10$^{5}$ km and by
the start of the {\it EUVE} observation (September 14), it had increased
to $\sim$2\arcmin~or $\sim$3 $\times$ 10$^{5}$~km. At this time, the 
cloud occupied a volume
of 10$^{31}$ cm$^3$ and the column density of dust had returned
to within a factor of 2 of its ``quiescent'' level. The cloud can 
hardly be discerned in the CCD image of September 16. {\it EUVE} continued
observing until September 19. 


For the gas images, only a CN shell was detected and only on September 
10. It had a radial extent about twice that of the dust, but with a relative
change in column density of 10 times less. However, CN is a daughter product 
which is believed to be partially released from the dust (A$'$Hearn et al. 
1995) and is therefore not representative of the gas outflow from the nucleus.
At these heliospheric distances, dust activity is driven largely by 
CO and H$_2$O emission (Weaver et al. 1997) to which the CCD 
images are insensitive. Biver et al. (1997) measured the outflow velocities 
of CO and OH (the main by-product of H$_2$O photo-dissociation) around 
the time of the LECS observation to be 0.5--0.7 km s$^{-1}$ i.e., at 
least 5 times that of the dust. Additionally, while the nucleus was 
out-gassing significant quantities ($\sim$3 $\times$ 10$^{29}$~s$^{-1}$) 
of both molecules, there is no evidence of a large gas outburst in the 
data of Biver et al. (1997). We caution however, that such an event 
could be missed since the lifetimes of CO and H$_2$O are of the order 
of days, and the data are infrequently sampled. 
 
\section{DISCUSSION}

The best-fit LECS bremsstrahlung temperature of $0.29\pm0.06$~keV 
is similar to that determined by \ROSAT\ for comets Hyakutake and 
Tsuchiya-Kiuchi (0.40~keV, Lisse et al. 1996; Dennerl, Englhauser, 
\& Tr\"umper 1996b). Figure~1 indicates that the bulk of the X-ray 
emission originates on the sunward side of the nucleus, also in 
agreement with previous observations. The emission may also show 
a characteristic elongation normal to the Sun-nucleus line. 
The similarity in properties between the LECS measurement of
comet Hale-Bopp and \ROSAT\ observations of Hyakutake and
Tsuchiya-Kiuchi further supports the view that the X-rays originate
from Hale-Bopp. The 0.1--2.0~keV luminosity of 
$4.8 \times 10^{16}$~erg~s$^{-1}$ is a factor 12 greater than observed 
from comet Hyakutake at a heliocentric distance of 1.0~AU 
and a factor 2.5 greater than from comet Tsuchiya-Kiuchi at a 
heliocentric distance of 1.4~AU. As we demonstrate below, the intense 
X-ray emission observed from comet Hale-Bopp by \sax\ is probably 
related to the large amount of dust present at the time.

The LECS spectrum is inconsistent with that expected from models which 
predict the bulk of emission to be in the form of lines, such as the solar 
fluorescent and the solar wind charge exchange models. 
In the case of comet Hyakutake, the plasma turbulence model of 
Bingham et al. (1997) predicts an  O~{\sc vii} 0.57~keV line with an 
intensity $\sim$10 times greater than the underlying
bremsstrahlung continuum (Kellett et al. 1997). For comet Hale-Bopp, 
the 95\% confidence upper limit to any such line emission is $<$16\% 
of the 0.1--2.0~keV continuum intensity or $<$10\% of the total luminosity. 
Models which involve perturbations of the solar wind 
or interplanetary magnetic field can also be excluded by contemporaneous 
{\it Geostationary Observational Environmental Satellite-8}
({\it GOES-8}) and {\it Solar Wind Experiment} ({\it SWE}) 
data which show that the solar X-ray and particle fluxes and the local 
magnetic field were stable at very low levels around the time of the 
LECS observation.

Krasnoplosky (1997) has shown that the attogram dust model can successfully
explain the measured fluxes from comet Hyakutake. Scaling the Hyakutake 
X-ray flux by the quiescent gas and dust production rates and heliospheric 
distances of Hyakutake and Hale-Bopp, the predicted Hale-Bopp X-ray flux is 
in good agreement with that measured by ${\it EUV}E$, under the assumption
that the X-ray emissivities are similar (Krasnopolsky et al. 1997b). 
Since X-ray intensity is expected to scale with the amount of dust, at 
least a factor of 3.4 (and more probably 8) more dust is required to 
account for the X-ray intensity 
observed by \sax, assuming the dust size distribution and solar X-ray 
flux remain the same. The dust production rate before the outburst is 
estimated to be 
$\sim$$4 \times 10^4$~kg~s$^{-1}$ and the ratio of dust to gas production 
$\sim$3 (Rauer, Arpigny, \& Boehnhardt 1997). The estimated dust
production rate during the outburst is at least 3 $\times$ 10$^5$~kg~s$^{-1}$, 
an increase by a factor of $>$7 compared to before the outburst, 
sufficient to account for the observed X-ray flux with the attogram dust model.
We cannot directly measure the corresponding change in the amount of gas 
surrounding the comet. However during previous outbursts the gas production 
rate increased by a factor of only 1--3 depending on the species 
(Weaver et al. 1997). Since the shell in Fig. 4 is thickening 
and expanding with time, its density is expected to vary as 
$\sim$v$^{-3}$, rather than v$^{-2}$ expected for a shell of 
constant thickness. Thus, assuming 
constant gas and dust velocities and spherical outflow, the gas density will 
decrease $\sim$200 times more rapidly than the dust, purely due to 
its higher expansion velocity.


By the start of the {\it EUVE} observation the dust density had 
decreased to a maximum of twice its quiescent level and the cloud 
had a size of $\sim$2\arcmin~or $\sim$3 $\times$ 10$^{5}$~km. By the 
end of the {\it EUVE} observation, the cloud filled most of the 
aperture providing an explanation for the source of extended 
emission reported in Mumma et al. (1997). The bulk of the dust outburst
occurred towards the North and North-West of the nucleus. It is interesting 
to note that the majority of the X-rays observed by {\it EUVE} 
also originated to the North of the nucleus (see Fig. 1
of Krasnopolsky et al. 1997a).  The reason that {\it EUVE} observed a  
lower flux than \sax\ is probably a combination of the continuation of the 
decay of the X-ray flux shown in Fig. 2 and poorer signal-to-noise 
caused by the spread of the signal over a much 
larger area of sky. In the absence of a more convincing explanation, 
we attribute 
this decay to dust fragmentation to sizes which are increasingly inefficient 
in producing X-rays.

Based on the following arguments we conclude that the observed X-rays are 
most likely produced in dust rather than gas. (1) the increase of dust 
produced during the outburst can explain the X-ray flux
measured by the LECS, (2) sub-micron grain sizes have been observed in the 
coma of Hale-Bopp (Williams et al. 1997), (3) the measured spectrum is 
inconsistent with
that predicted by gaseous models but consistent with that 
predicted by the attogram dust model, (4) the spatial distribution of X-rays 
measured by {\it EUVE} is similar to that of the dust from the outburst, and 
(5) X-rays were not observed from comet Bradfield 1979X - a highly
gaseous but the least dusty comet observed (Hudson et al. 1981).

\acknowledgements

{\it GOES-8} data were kindly provided by D. Wilkinson of the 
NOAA/National Geophysical Data Center and {\it SWE} data by K.W. Ogilvie 
(NASA/GSFC), J.T. Steinberg (MIT), and A.J. Lazarus (MIT). K. Dennerl 
is thanked for many useful conversations. A. Orr and T. Oosterbroek 
acknowledge ESA Research Fellowships. The \sax\ satellite is a joint 
Italian and Dutch programme.

\clearpage
\begin{table*}
\begin{center}
\centerline{COMETARY X-RAY EMISSION MODELS}
\vspace{0.5cm}
\begin{tabular}{lll} 
\tableline
\tableline
Emission Mechanism & $\epsilon^a$ & Ref.  \\
\tableline
Ionization by solar wind protons           & 10$^{-6}$      & 1\\
Current sheets in the solar wind           & \llap{$>$}10$^{-6}$ & 2\\ 
High velocity dust-dust collisions         & 10$^{-5}$      & 1\\
X-ray scattering in coma                   & 10$^{-3}$      & 1\\
Solar X-ray fluorescence                   & 10$^{-3}$      & 1\\
Bremsstrahlung from cometary electrons     & 10$^{-2}$      & 3\\
accelerated by plasma turbulence           &                   \\  
Charge exchange of heavy solar wind ions   & 1              & 4\\
with neutral cometary molecules            &                   \\ 
Solar X-ray scattering, fluorescence and   & 2              & 1\\
bremsstrahlung in attogram dust particles  &                   \\
\tableline
\end{tabular}
\end{center}
\tablenotetext{a}{The ratio of predicted to measured fluxes for comet Hyakutake.}
\tablenotetext{1}{Krasnopolsky 1997; $^2$Brandt, Lisse, \& Yi 
1997; 
$^3$Bingham et al. 1997; $^4$H\"aberli et al. 1997.}
\tablenum{1}
\end{table*}

\clearpage
\begin{figure*}
\centerline{\psfig{figure=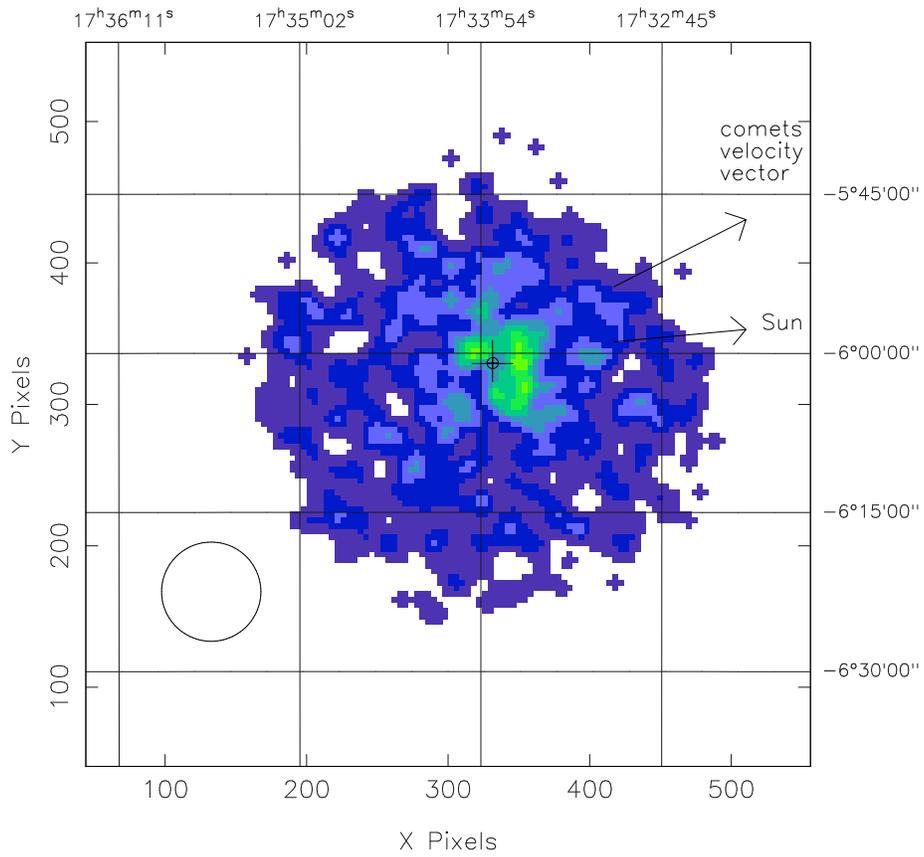,width=15cm,angle=-90}}
\caption{
A 0.1--2.0~keV image of the sky containing 
comet Hale-Bopp. The image has been corrected for the motion of the comet 
($\sim$ 17$''$ hr$^{-1}$). For comparison, the FWHM of the PSF at the mean 
energy of the source is indicated by the circle in the lower left hand 
corner. The position of the nucleus is indicated by the small circle and 
cross. The position of the Sun and the comet's velocity vector are indicated.
}
\end{figure*}
\vfill

\clearpage
\begin{figure*}
\centerline{\psfig{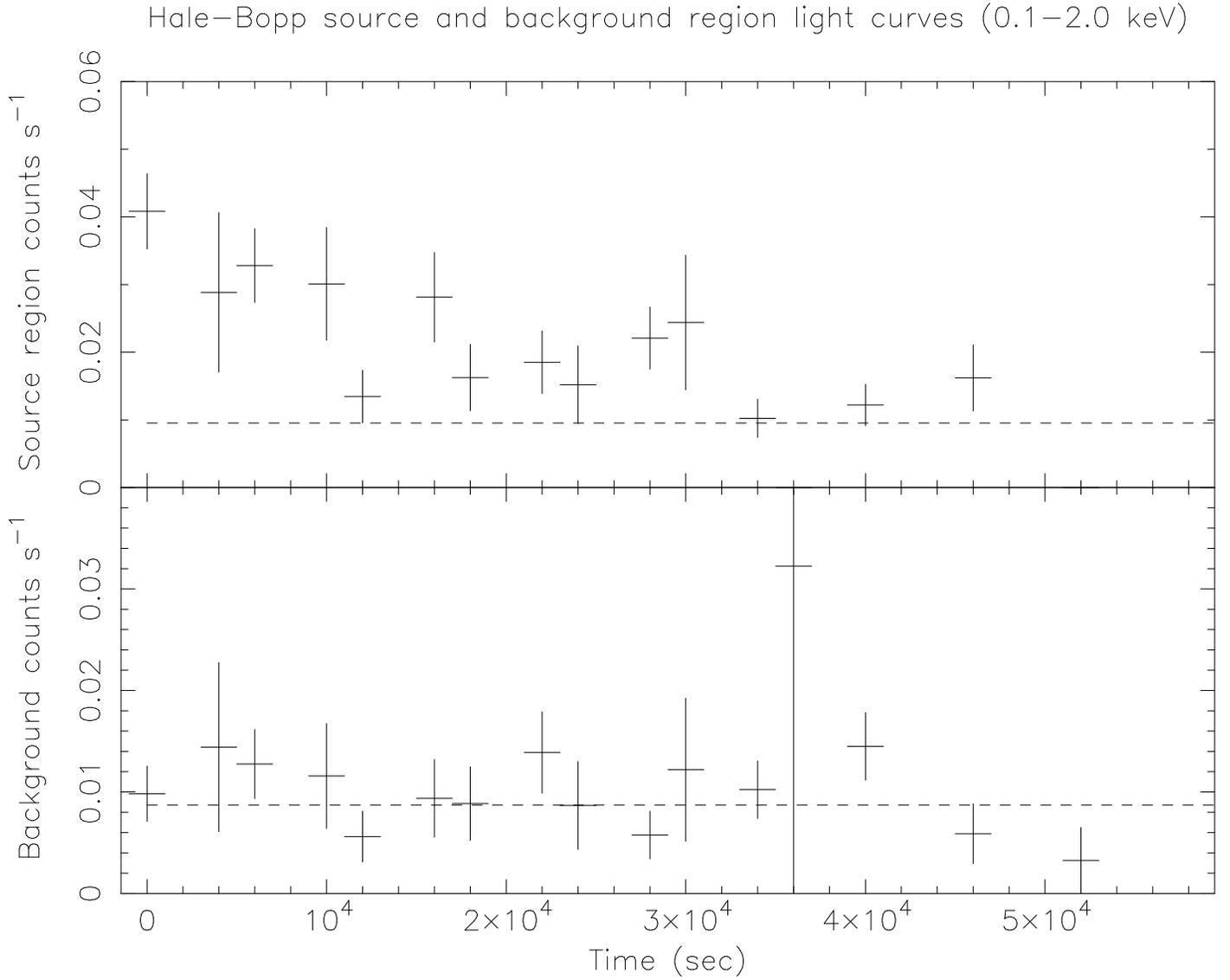}}
\caption{
The 0.1--2.0~keV light curves of source and background regions. 
The mean count rate in the background region is shown by the dotted line 
in the lower panel. Based on this, the predicted background level at the 
position of the comet (corrected for the difference in vignetting) is 
indicated by the dotted line in the upper panel.}
\end{figure*}
\vfill

\clearpage
\begin{figure*}
\centerline{\psfig{figure=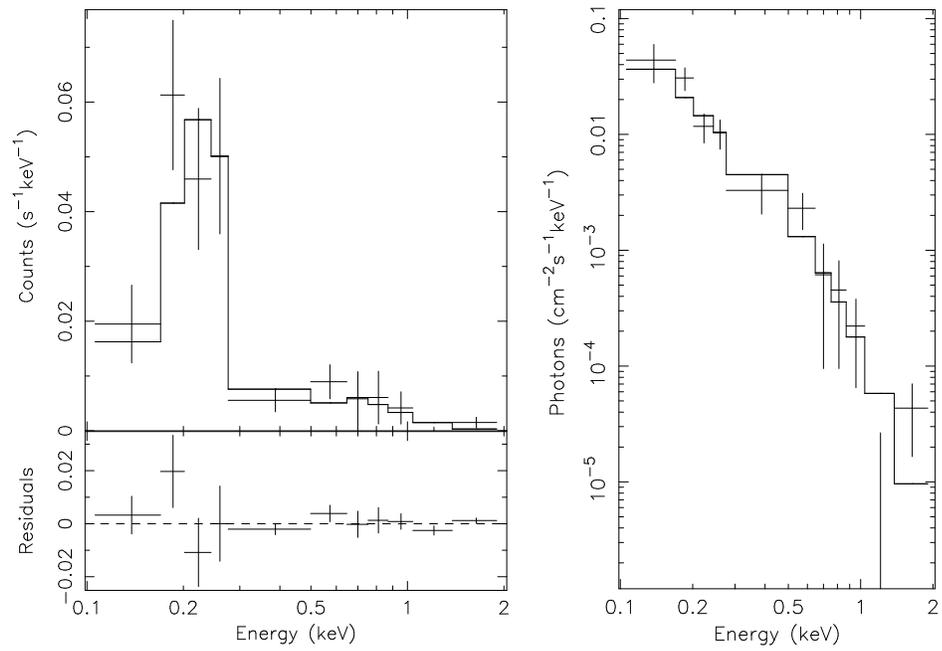,width=15.0cm,angle=-90}}
\caption[]{
The observed LECS Hale-Bopp spectrum and best-fit thermal 
bremsstrahlung model (kT = 0.29~keV). The lower left panel shows 
the residuals. The inferred photon spectrum and model prediction 
are shown in the right panel.}
\end{figure*}
\vfill

\clearpage
\begin{figure*}
\centerline{\epsfig{figure=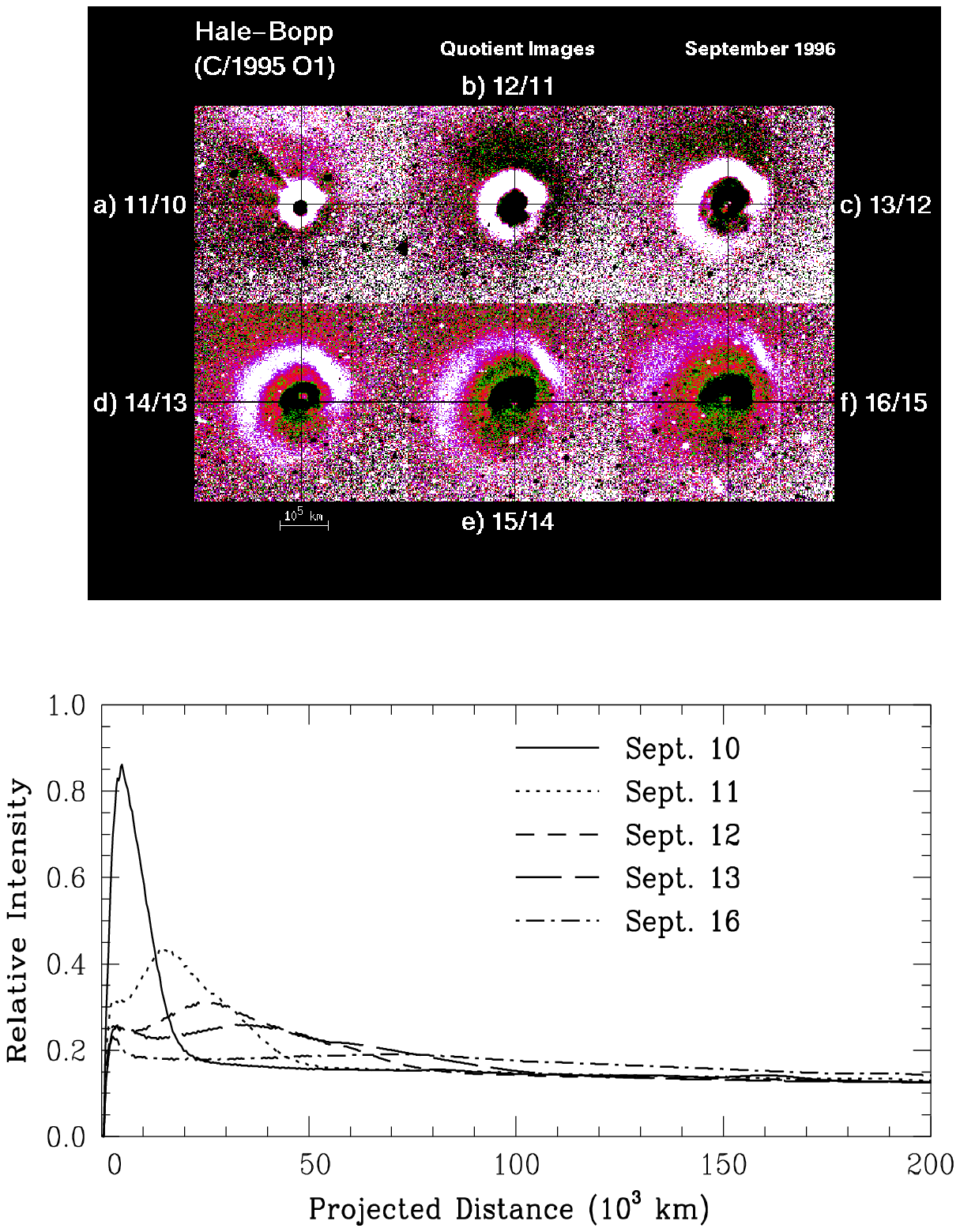,width=12.0cm,angle=0}}
\caption[]{
The temporal evolution of the dust shell 
observed in Hale-Bopp between
1996 September 10 and 17. Each image is the ratio between an image and that
obtained the previous day and thus highlights changes between 
consecutive days. 
The images have the same orientation as Fig.~1. The intensity coding of the 
quotient images has been modified to enhance the resulting shell in each case. 
White indicates a large change and black no change. Note, as the shell is 
expanding with time, each image is divided by one with a smaller and more 
intense shell. This artificially increases the dimension of expanding shell 
structures and hence the apparent expansion velocity of the shell.
Lower panel:  Differential dust band ($\lambda$=4467~\AA) radial intensity 
profiles obtained by differentiating
the intensities in concentric circular apertures centered on the maximum 
intensities in the dust coma. A clear evolution of the dust shell is evident. 
The ratio of the peak intensity on September 10 and the quiescent
value (as measured 2~$\times$ 10$^5$ km from the 
comet on 1996 September 10) implies a change in the dust column density 
by at least a factor a 7.}
\end{figure*}

\end{document}